# Intelligent Agents: A Physics Education Opportunity in Latin-America

**D. Sánchez-Guzmán**, **César Mora**, **R. García-Salcedo**
*Centro de Investigación en Ciencia Aplicada y Tecnología Avanzada, Instituto Politécnico Nacional, Legaria No.694, col. Irrigación, Miguel Hidalgo, CP 11500, México D. F.*

**E-mail:** dsanchez@ipn.mx; cmoral@ipn.mx; rigarcias@ipn.mx

**Abstract**
Intelligent Agents are being applied in a wide range of processes and everyday applications. Their development is not new, in recent years they have had an increased attention and design; like learning and mentoring tools. In this paper we discuss the definition of what an intelligent agent is; how they are applied; how they look like; recent implementations of agents; agents as support in the learning process; their state in Latin-American countries and future developments and trends that will permit a better communication between people and agents.

Los Agentes Inteligentes en la actualidad se implementan en una gran cantidad de procesos y aplicaciones. El desarrollo de estos no es nuevo, en años recientes se ha incrementado el desarrollo y diseño como herramientas de tutoría. En el presente artículo se presenta la definición de lo que es un agente inteligente, sus aplicaciones más frecuentes, el soporte que estos brindan en el proceso enseñanza aprendizaje; así como el estado actual en países Latinoamericanos. Se presenta también el desarrollo y las tendencias de estos agentes que permitirán una mejor comunicación entre las personas y otros agentes.

## I. INTRODUCTION

Current technologies have been applied in many social and personal processes. Now computers are being implemented in many activities, like personal web pages, on-line systems for government, factory systems, education systems, etc. Internet have evolve from a simple text-based scheme to a multimedia-based scheme; before we had Hyper Text Transfer Protocol (HTTP) as the default protocol and Hyper Text Mark-up Language (HTML) as the displaying tool in a browser, like Internet Explorer or Netscape, both to transfer information between enterprises, people and automated systems, all this happened using the World Wide Web (WWW). It did not pass a long time until that technology becomes obsolete, it was necessary to create and enrich the Web content, not only by displaying more information and images; also it was necessary to give to the information order and coherence, because we were falling in an huge accumulation of information process, and the content was not classified and ordered correctly [1].

Then their appeared technologies for solve many problems of order and content that Internet was suffering; so their appeared technologies like eXtensible Mark-up Language (XML) as a standard, Personal Home Pages (PHP) as an open project, Active Server Pages (ASP) from Microsoft Corp., Java Server Pages (JSP) from Java Technologies by Sun Microsystems, and others. These technologies help us to create web content with more order and coherence, and facilitate the manipulation of information. Also have appeared tools for managing, controlling and searching inside all these amounts of information, then we had the development of Data Bases (DB) and Data Bases Management Systems (DBMS) like MySQL or PosgresSQL, the exponential growth of information had accumulated as many information as the entire life of each person in the world, nowadays we can talk that information has an amount of PetaBytes (PB) or ExaBytes (EB) that is many information for administrating, maintaining and searching; for controlling such amount of information there were developed software robots (now know as 'softbot') and techniques for controlling all that information, the result of these development where Data Warehousing Systems, Data Mining Systems and others, many of them apply Artificial Intelligence (AI) techniques.

The merge of technologies, techniques and methodologies have created solutions that were directly applied on social processes, this permitted to expand solutions in a quickly way. As an addition, the boom of telecommunication systems and technologies and their integration have permitted people to be communicating to each other all over the world. As an example two people can talk each other using an Internet chat or video conference from different and very distant cities or countries [2]. Cellular networks are another example, and now we have the concept of Personal Area Network (PAN). The definition that integrates all the described above is defined as Information and Communication Technologies (ICT), from now to the rest of this paper we refer only like ICT to all solutions and definitions of applied technology.

The aim of this paper is to present actual nature of Intelligent Agents; in Section II we present what they are, how they look like and how they work. In Section III we present the Teaching Agent as an application in education and its behavior. In Section IV we present the Intelligent Agents Applied to Physics Education, an example of how intelligent agents are applied to Physics Education. In Section V are shown the Future Trends of Intelligent Agents. Finally, in Section VI we present the conclusions obtained about intelligent agents and a proposal of applications and future research.

## II. INTELLIGENT AGENTS

During the past five decades there have been developed software tools that would help people to facilitate their work. They have evolved from an ambiguous and esoteric interface only permitted to be manipulated by an expert to a friendly child-use interface; this let people to interact with computational systems and manipulated them very easily.

The development of programming paradigms like Object Oriented, Event Oriented, etc., has permitted to create a wide range of applications. Areas like Artificial Intelligence are being employed to create Intelligent Agents; this can be applied in the education processes to facilitate the students in their learning and development of skills, and they are focusing on what we know like hard subjects as Math and Physics. The rest of this section we will explain what an intelligent agent is and how it should look like.

### A. What is an Intelligent Agent?

The idea of agents is not new. For over five decades researchers from a wide range of fields have studied problems that demonstrate some type of agent behavior [3]. According to Pattie Maes of MIT, intelligent agents "*are computational systems that inhabit some complex, dynamic environment, sense and act autonomously in this environment, and by doing so realize a set of goals or tasks that they are designed for.*" [4] These autonomous agents can have different appearances depending on the environment they exist in. If the environment is the real physical environment, then the agent takes the form of an autonomous robot. There are also 2D or 3D animated agents which inhabit simulated physical environments and 'knowbots.' which are software agents or interface agents which inhabit the digital world of computers and computer networks [4].

There is a range of characteristics of the agents derived from various descriptions. They are considered autonomous, having the ability to '*take the initiative*' and can formulate their own goals and acting in order to satisfy them, with other agents and with humans [5]. The communication may be in the form of text commands or more recently speech and even gestures [4]. The ability to learn, mentioned later in this paper is also seen as an intelligent agent characteristic, as is their ability to perceive their environment, and respond to changes that occur [5].

### B. How should an intelligent agent look?

Overall, the agent is preferably portrayed with graphic or iconic representations rather than realistic animations or video, although more realistic animations have been explored such as Microsoft's conversational parrot '*Peedy'* [6] '*Herman the Bug*' in the Design-a-Plant learning environment. But Shneiderman [7] suggests that user interfaces should not attempt to mimic human interaction but maintain a "neutral" status.

A good compromise may be the use of different faces, which Maes [8] describes in an email agent application. Specifically, the agent has different faces to indicate what the agent is doing: thinking, working, suggesting, unsure, pleased, confused. In terms of the value of agents exhibiting life-like expressions, Lester and Stone [9] and Koda and Maes [10] suggest that such agents have greater motivational impact.

### C. How an Intelligent Agent Works?

The social relationship between the learner and agent is further defined to be a pedagogical relationship, where the intelligent agent (in theory) can monitor and evaluate the timing and implementation of teaching interventions (e.g. provide help, feedback). And, on the other hand, this relationship can be described as a cognitive apprenticeship, where the student improves his performance while working with the most expert performer: the intelligent agent.

In support of this feature, Collins and Brown [11] suggest environments that include modeling and coaching of formative skills are the best for the learning in the students. A favorable situation with intelligent agents is that while the student gains expertise, the agent would fade and allow to student take the initiative for his/her learning. An important consideration in terms of feedback is that the intelligent agent should not provide too many insights and thereby annoy the student. Furthermore, as Negroponte [12] suggests, the human act of winking can connote a lot of information to others simply in the *lack* of information

This sort of familiarity is needed for the pedagogical agent to avoid relentless or explicitness. To address this issue, part of the pedagogical task should include the monitoring of the timing and implementation of the advisements. Using the principle of minimal help as the default in the intelligent agent, the student could select a feedback option depending on the amount of structure, interaction, and feedback it desire when he/she solve problem and exercises There are significant difficulties in developing pedagogical expertise in an intelligent agent. As McArthur, Lewis and Bishay [13] state, the pedagogical component of intelligent systems receives relatively little mention with current systems demonstrating little pedagogical expertise.

As they suggest [13], most intelligent tutoring systems are constrained to a single method of teaching and learning, while truly expert human tutors can adopt different methods. Dillenbourg, Mendelsohn and Schneider proposed an intelligent agent called the ETOILE system [8], however, it consists of multiple pedagogical agents taking different approaches according to educational psychology principles. The five teaching agents in the ETOILE system are named Skinner, Bloom, Vygotsky, Piaget, and Papert where Skinner works step by step, Bloom takes larger steps but with control of mastery, Vygotsky is based on participation, Piaget intervenes only to point out problems, and Papert does not interrupt the learner. Further, their system includes a "coach" agent that manages the roles of the pedagogical agents and controls the interactions among all of them and the learner.

Another system developed is Baylor's MIMIC (Multiple Intelligent Mentors Instructing Collaboratively) system [14], which consists in three pedagogical agents (from three different theoretical perspectives) to teach instructional design to pre-service teachers. Specifically, the three pedagogical agents take three different perspectives of instructional design: the *instructional systems design* pedagogical agent; the *alternative views of instruction* pedagogical agent; and, the *teaching as an art* pedagogical agent. In this system, each agent interacts with both the learner and the others pedagogical agents to provide a unique and holistic learning environment for instructional design.

The Andes Physics Tutor is another kind of agent development. Andes is an intelligent homework helper for physics. That is, it replaces the pencil and paper that students would ordinarily use to solve physics homework problems. Students draw diagrams, enter equations and define variables with the same freedom that they have when using paper. Yet, unlike a piece of paper, Andes tells students whether their entry is correct by turning it red or green, and Andes will give principle-based hints when is asked [15]. It is free, courtesy of funding from the Cognitive Science Program of the Office of Naval Research and the National Science Foundation. Andes was developed at the University of Pittsburgh and the United States Naval Academy by VanLehn, K. and colleges, today the project is been developed at Carnegie Mellon University.

Cognitive Tutor is an intelligent agent, from Carnegie Learning, a ten-year-old company that started activities at Carnegie Mellon University, designed to give students individualized instruction when personal attention is scarce. Although such intelligent tutoring systems have their share of skeptics, students at schools that use them have not only improved their performance in math but now enjoy to learn a subject they once loathed [10]. The artificial intelligence built into the Carnegie Learning program helps set it apart. Not only does the program present drills according to a student's weaknesses, but it watches the work step by step, detecting where the student stumbles, and in this form the agent takes part when is necessary.

uMind it is another intelligent agent software, through a decade of research, Frasson has found a way to integrate artificial intelligence (AI) and Advanced Learning Techniques (ALT) into computer-based education. He says that uMind has taken eLearning beyond the pages of static text fed onto a computer screen [10]. Instead, uMind's teaching tool adjusts to the strengths and weaknesses of each student. "*Intelligence means adapting to the learner and understanding the capability of the learner,*" said Frasson, founder and president of uMind. "*In any exam, you have a lot of learners who are able to succeed. We have found all the mechanisms (that help students succeed) and we have applied them to eLearning.*"

## III. THE TEACHING AGENT

We are interested in the teaching agent that must have a combination of skills. In some cases, it will be required to be a '*knowbot*' providing the interface between learner and the huge amount of information. In other occasions, the agent needs to 'inhabit' the learning environment of the students, mentoring them through the material they need to learn. In this situation the agent has to be aware of its environment. It has to decide what to do next in order to provide information or teach the learner the next step in their learning. Prior to 1982, teaching by computer was relegated to the field commonly known as computer-aided instruction (CAI) [16].

As more versatile software emerges, the boundary between basic computer aided instruction and more intelligent systems has been blurred. In many of the current CAI packages, the software tries to match the level of training to the learners skills. It has to assess the learner skill and adjust the learning to that level. Compared with artificial intelligent (agent) systems the user interaction is still too restrictive, 'limiting the student expressiveness and thereby limiting the ability of the tutor diagnostic mechanisms' [16]. The CAI software has a limit based on generalized responses to questions.

Broadly defined, an intelligent tutoring system is educational software containing an artificial intelligence component. The software tracks the students work, tailoring feedback and hints along the way. By collecting information on a particular student performance, the software can make inferences about strengths and weaknesses, and can suggest additional work [10].

As we can see the intelligent agent has a different approach. Agent based-education opens up whole new and exciting areas of possibility because the agent does not simply teach. The intelligent agent is more than just the teacher, it is also the learner. It adapts to the changing needs of the student, their environment and learning style. The agent builds up a character map of the learner in the same way that in classroom teacher does. The technology makes feasible to have a personal computer assistant that keeps building a database of everything you do, including continuous real-time videos. "*Soon, perhaps, it could also record vast stores of information about the state of your brain while you did things. Perhaps, in your wristwatch, or in a permanently implanted nanotechnological computer module*" [17].

## IV. INTELLIGENT AGENTS APPLIED TO PHYSICS EDUCATION

There are a wide variety of applications that have been applied to educational processes; we have mentioned ETOILE, MIMIC, Andes, Cognitive Tutor and uMind as tools for educational teaching, in this topic we will focus in tools applied to Physics Education, to be more precisely we explain the Andes Physics Tutor and Cognitive Learning, how they interact with students, with teachers and in the learning-teaching process.

The principal reason to use Andes is that it helps students *learn more* from solving physics problems. The second reason is that Andes acts as a *homework grading service*, thus reducing the burden on the course instructors. Andes produce learning gains in two ways: First, because Andes can grade every student solution to every homework problem, Andes makes it feasible for instructors to include homework problem solving as part of the students course grade, which may motivate students to actually do their homework. This can cause increases in learning [18].

The second way that Andes produces learning is by encouraging students to do semantically well-formed problem solving that leads to a correct solution. As mentioned earlier, Andes *requires* certain practices (e.g., dimensional numbers in equations must have proper units) and it *encourages* others by giving extra points. Moreover, its step-based feedback and hints virtually guarantee that all students can solve all problems correctly. In order to test whether these features increase learning, it was conducted a rigorous, four-year experiment at the United States Naval Academy [19], which will be described next.

Some important considerations about Andes: a) Andes currently runs only under the Windows operating system; b) Using Andes user interface requires learning some non-obvious details, such as how to draw a zero-length vector. Both students and instructors should study the 14 minute user interface training video before trying to solve Andes problems; c) Andes does not replace a textbook or an instructor. It is simply a large set of problems. It is intended to be used as part of a college, high school or distance learning course; d) Andes may be used with any introductory physics textbook that teaches problem solving using algebra and trigonometry [19]. Andes problems currently do not require writing integral or differential equations, but neither does many calculus-based physics textbooks.

Andes has problems for all the major chapters in introductory physics textbooks, except for the chapters on thermodynamics and modern physics, which are still under development. However, Andes currently does not have very many problems per chapter. It is constantly adding problems, and welcome suggestions for which ones to add.

Andes is not a perfect tutor. Although it seldom crashes and it is almost always accurate when labeling a step correct versus incorrect, students sometimes do not understand its hints. This is understandable if you pick up a student's partially solved physics problem, try to decide what the student is doing and what would be a good next step, and then give a hint on that step. Even an expert hints fail sometimes. Moreover, students with particularly deep misconceptions may not understand some of Andes' hints. Thus, the instructor should still expect and encourage students to ask questions about their homework. Some students may also need help with the user interface, at least initially [19].

Instructors may occasionally be baffled by Andes' behavior. For instance, one instructor could not figure out why Andes would not accept a solution to a problem that the instructor knew could be solved by conservation of energy. It turned out that the problem was taken from the Newton's law problem set, where all energy-based solutions are prohibited. The same problem was available in another problem set, where all solutions were allowed. During the first year of use, an instructor can expect to access the Andes FAQ list or to contact the Andes team several times in order to learn about hidden "features" such as this one and to report suspected bugs.

In short, although Andes does increase student learning, and it currently costs no money to use it, it does cost some time from both instructors and students as they learn how to use it. Although we do not have exact figures, it appears to take about an hour to become comfortable with Andes initially, and perhaps ten minutes every month thereafter to learn its hidden *"features"* [19].

On the other hand, Andes probably saves more time than it requires. On US Naval Academy questionnaires, most students report that, in their opinion, they spend less time doing their homework on Andes than they would if they had to do the same problems on paper. Instructors probably save time by not having to grade papers or supervise the grading of papers. Physics departments that have replaced human graders with homework grading services have reported considerable cost savings.

Cognitive Learning is the other tutoring tool that we will examine. The Carnegie Learning software is one of the few computer-based instructional programs on the market. The e-learning industry is huge, with hundreds of companies selling various forms of e-learning software.

But experts dismiss much of what is available as basic computer-aided instruction, offering little more than a question-and-answer format [9]. The artificial intelligence built into the Carnegie Learning program helps set it apart. Not only does the program present drills according to a student's weaknesses, but it watches the work step by step, detecting where the student stumbles, and chimes in when necessary.

"*People solve problems in different ways",* said Ken Koedinger, a professor of human-computer interaction and psychology at Carnegie Mellon, and a co-founder of Carnegie Learning. "*So a student who solves a problem one way gets a different hint than a student who solves in a different way. You don't get that with computer-aided instruction systems*" [9]. The software does not replace classroom teaching. At many schools the program is used at a different time from math class, in a separate computer lab.

But intelligent tutoring systems have their skeptics. One of the main criticisms is that the tutor does not really allow students to learn from their mistakes because the program vigorously guides them toward the right answer [9]. This skepticism can be for many reasons; one considerable is the level that technology can support the applications, that is, the capacity of calculating or processing task of a computer in a fast way.

As we can see, there are pros and cons with tutoring tools; one important fact mentioned by researches in education is that these can not supply the teacher instruction like is showed by many research in this area. It is necessary to develop more tools and continue with the research on Artificial Intelligence topics, this will help to create an inclusive education system for all over the world. Interdisciplinary research groups are being formed to study the case of tutoring systems, it is important that psychologists, pedagogues, institutions, computer science scientists, teachers and more important students work jointly to solve many education topics in learning education.

A point to observe is that these tools and others are well implement in U. S. A. and Europe. Countries in Latin-America are in an incipient development, this is an opportunity to researches in education and the implementation of Information Technologies (IT).

## V. FUTURE TRENDS OF INTELLIGENT AGENTS

Current technologies on Intelligent Agents permit to society perform some task that have helped enormously in many processes and situations, and have liberated people from calculating tedious tasks; also they have been an extra helper that let people to take decisions. The continuous development in agents will permit more interacting tools, give them more reality and an interface human-like, this will permit experiment a new age in education and training.

Instead of agents are being developed to collaborate between other agents, humans and network tools. Interfaces must change, we have been working with a mouse-keyboard-based interface for over three decades, to achieve the development of a more human-like interface it is necessary to develop and imagine new types of interfaces. The increase of computation in machines will permit make agents more rich and give them more intelligence and response capacity.

In the education field agents must be adequate to teaching people and must work according to theories at the present in cognitive research; agents tend to accomplish they objectives of what they were designed like autonomous, work in complex environment, interact with other agents, have a complex and easy to use interface, etc.

## VI. CONCLUSION

Intelligent Agents are being applied in many disciplines and in a variety of scientific research areas. It is important to know that the development of these tools is increasing, and that it let us to implement them in a wide range of applications. In the education field intelligent agents are going to help students to improve their skills, serving as tutor, this solution of agents is not new, but has not been widely development because a lot of factors like computational processing and also the interfaces have not been developed to accomplish the communication between people and agents, also the integration of cognitive theories and computational theories did not permit a growth development.

In Latin-American countries Intelligent Agents are not very implemented as solutions this could be because the most programs and curricula do not include this kind of tools, also infrastructure of many universities, institutes and schools do not have the minimum requirements of equipment. This is an opportunity to develop and research on Intelligent Agents thru these countries and adapt them to particularities of Latin-American people. Students, teachers and researchers have lot of work to do in developing agents applied in education and learning processes.

## ACKNOWLEDGEMENTS

This work was supported by COFAA and EDI IPN grants, SIP-20090739, SIP-20090440 and CONACYT-91335 Research Projects